\documentclass[10pt,twoside]{classe-cmb}
\usepackage{amssymb,amsbsy,amsmath,amsfonts,amssymb,amscd}
\usepackage{referj}
\usepackage{latexsym,euscript,exscale,epic,eepic,epsfig}
\usepackage[english,francais]{babel}
\usepackage{times}
\newtheorem{e-proposition}[theorem]{Proposition}

\newtheorem{e-definition}[theorem]{Definition\rm}

\ComParit{S1296-2147}
\PIT{FLA}
\PXHY{????}
\Add{?}
\Volume{0}
\Year{2003}
\FirstPage{1}
\LastPage{??}
\AuteurCourant{A. Blanchard}
\TitreCourant{Cosmological parameters} %less than 60 characters
\Journal
\Rubrique{Rubrique}{Heading}
\SousRubrique{Sous-rubrique}{Sub-Heading}
\PresentePar{First name}{NAME}
\Recu{blank}{}{}
\keywords{keyword~1~/ keyword~2~/ etc.}
\begin{document}
\selectlanguage{english}
\TitleOfDossier{The Cosmic Microwave Background: \\present status and
  cosmological perspectives}
\title{%
Cosmological implications from the observed properties of
CMB.
}
\author{%
Alain Blanchard~$^{\text{a}}$, 
James Bartlett~$^{\text{b}}$\&
Marian Douspis~$^{\text{c}}$
}
\address{%
\begin{itemize}\labelsep=2mm\leftskip=-5mm
\item[$^{\text{a}}$]
Laboratoire d'Astrophysique de l'Observatoire Midi-Pyr{\'e}n{\'e}es,
14 Avenue E.~Belin, F--31400 Toulouse, France \\
E-mail: Alain.Blanchard@ast.obs-mip.fr
\item[$^{\text{b}}$]
APC, Universit\'e Paris 7, Paris, France\\
E-mail: bartlett@cdf.in2p3.fr
\item[$^{\text{c}}$]
Astrophysics, D. Wylikinson bd., Keble Road, Oxford OX1 3RH,  United Kingdom\\
E-mail: douspis@astro.ox.ac.uk
\end{itemize}
}
\maketitle
\thispagestyle{empty}
%%%%%%%%%%%%%%%%%%%%%%%%%%%%%%%%%%%%%%%%%%%%%%%%%%%%%%%%%%%%
%%%  Abstract  %%%
%%%%%%%%%%%%%%%%%%
\begin{Abstract}{%
    L'ensemble des observations du fond cosmologique repr\'esentent
    une source d'informations remarquable pour la cosmologie moderne.
    Non seulement elle conforte le mod\`ele du ``Big Bang'', mais
    pr\'ecise notablement le cadre dans lequel les structures de
    l'univers se sont form\'ees. Mais ce qui est sans doute le plus
    fascinant est que le fond cosmologique est la voie privil\'egi\'ee
    d'acc\`es \`a la physique des tr\`es hautes \'energies qui
    r\'egnaient dans les tous premiers instants de l'univers et qui
    pourraient demeurer \`a jamais inaccessibles de fa\c con directe
    aux exp\'eriences de laboratoire.  Les futures exp\'eriences, dont
    Planck en particulier, sont donc une porte ouverte sur la physique
    du troisi\`eme mill\'enaire.\\
}\end{Abstract}

%\medskip\centerline{\rule{2cm}{0.2mm}}\medskip

\begin{Abstract}{%
    Observations of the cosmic microwave background represent a
    remarkable source of information for modern cosmology.  Besides
    providing impressive support for the Big Bang model itself, they
    quantify the overall framework, or background, for the formation
    of large scale structure.  Most exciting, however, is the
    potential access these observations give to the first moments of
    cosmic history and to the physics reigning at such exceptionally
    high energies, which will remain beyond the reach of the
    laboratory in any foreseeable future.  Upcoming experiments, such
    as the Planck mission, thus offer a window onto the Physics of the
    Third Millennium. 
}\end{Abstract}

\par\medskip\centerline{\rule{2cm}{0.2mm}}\medskip

\setcounter{section}{0}
\selectlanguage{english}
%%%%%%%%%%%%%%%%%%%%%%%%%%%%%%%%%%%%%%%%%%%%%%%%%%%%%%%%%%%%
%%%  Main text (in English)  %%%
%%%%%%%%%%%%%%%%%%%%%%%%%%%%%%%%

\section{Introduction}

The program of modern cosmology was born with Lema\^{\i}tre's 1927
paper in which he proposed a cosmological model primarily motivated by
the desire of accounting for what he believed to be the two
astronomical facts of major significance for the description of the
Universe: its non--zero matter content and the apparent recession of
galaxies that he interpreted as a direct evidence for the expansion of
the Universe. A few years later, after the clear evidence for an
expanding universe obtained by Hubble, Lema\^{\i}tre initiated a
program whose basic questions still represent fundamental lines of
research in modern Cosmology: the very early history of the Universe,
including the nature of the initial singularity and its connection to
quantum mechanics, and the question of the history of structure
formation.  During the rest of the XXth century, cosmology underwent
 remarkable progress, by the continuation of this confrontation of
some of the most recent, often regarded as the most exotic, theories
in physics, with hard  astronomical data.  The determination of the
values of the cosmological parameters, with a moderate error, has
naturally always being one of the central goals of cosmology, although
the strength of efforts in this direction has varied over time.

The concept of inflation (A.~Guth, 1981), introduced more than twenty
years ago, revolutionized the field, pointing out how Cosmology
contained deep connections between high energy physics and some
astronomical observations.  Moreover, Inflation suggests that the actual
value of several quantities could have a {\em physical origin}, rather
than being just constants that have to be determined.  In addition,
the need for a {\em physical origin} of the fluctuations that seed
structure formation reinforced the link between the question of
large--scale structure and the physics of the Big Bang. It was
recognised during the last twenty years that the properties of
large--scale structure, as revealed by the galaxy distribution, was a
potential source of key information for understanding the physics that
occurred during the very first instants of the universe. For these
reasons, the determination of cosmological parameters has become a
scientific program which significance goes far beyond the question of
establishing the numerical values of the few parameters describing the
universe within the framework of general relativity.

Establishing the precise abundance of light elements, which requires
the modelling of the chemical evolution of galaxies and therefore the
precise understanding of the physical process occurring in stellar
interiors of stars, has been a fundamental test of the Big Bang during
its first minutes, and is a good example of connections between modern
cosmology, some fundamental physics (nuclear physics in this example)
and classical astrophysics. There is now good convergence of data to a
rather restricted range of possible values for the baryonic content 
(Charbonnel 2002) of
the Universe: $$\Omega_b \sim 0.022 h^{-2} \pm 10\%$$ This convergence
makes Big Bang nucleosynthesis one of the piller of modern cosmology.

The discovery of the CMB provided the third fundamental piller on
which the standard Big Bang is build.  The verification of its
remarkable black body spectrum by FIRAS/COBE represents the essential
last achievement of the "classical cosmology" program, allowing a
reliable description of the major points of the history of the
universe between the first billionth second and the present epoch.
However the discovery by DMR/COBE of the fluctuations of the microwave
sky has brought an essential observational fact that requires physical
explanation beyond the physics well--established in laboratories.
Whether inflation is the correct explanation of the origin of the
fluctuations in the observed spectrum of the angular fluctuations in
the microwave sky is still a matter of debate, although it is
remarkable that this theory proposed more than twenty years ago has
passed remarkably well several observational tests.  But the need for
new physics is increasingly evident. It has also become clear that
Cosmology will provide a test bed for this high energy physics that
may well remain unattainable otherwise.  In this respect, the observed
properties of the CMB fluctuations appear as a remarkably clean tool
for investigating early high energy physics.  In fact, the possibility
of constraining some parameters to the percent level with Planck, an
extraordinary challenge, naturally leads to the idea of "high
precision cosmology" in a scientific domain where order of magnitudes
were the only realistic perspective few years ago!

\section{Why CMB does tell us something on cosmological parameters?}

In the standard scenario of structure formation, the present
distribution of matter results from the gravitational amplification of
initially small perturbations of the matter density field. As the
temperature of the universe goes down, the initial hot plasma will
eventually recombine in neutral gas, suddenly leaving the universe
essentially transparent. Therefore, observing the cosmic microwave
background offers a direct image of the universe at this epoch, some
400 000 years after the Big Bang (It is instructive to recall that
this image is and will remain the most distant picture of the universe
that light could ever reveal!).  The physical conditions presiding at
this epoch are well known and easy to describe (the density of matter
at this epoch is still lower than the best vacuum one can obtain in
laboratories!), the amplitude of the fluctuations being in the linear
regime. Therefore, the calculation of the angular spectrum of
temperature anisotropies, $C_\ell$'s, resulting from a given initial
matter fluctuation power spectrum $P(k)$, whose statistics is
specified, is relatively straightforward even if it could be quite
elaborate on the technical side. Qualitatively, fluctuations behave
like waves in a viscous media -- they oscillate and the amplitude
decreases with time. This specific oscillating regime starts when the
wavelength become smaller than the horizon. Therefore each wavelength
starts oscillating with a fixed value of the initial phase but at an
epoch varying with the wavelength. This oscillating regime stops
rather brutally when the universe become transparent at recombination.
The specific amplitude of the wave at that time depends on this phase
(as well as on the detailed composition: baryonic and non-baryonic
matter) and imply a specific pattern at some spatial wavelengths and
its harmonics, which appear as successive peaks in the $C_\ell$'s
curve.  These peaks are the angular equivalent of the specific spatial
wavelengths. Therefore their numerical values depends on the angular
distance to this surface corresponding to this epoch and involves
various cosmological parameters. This allows one to understand why the
$C_\ell$ curve depends on the characteristics of the spectrum of the
initial fluctuations, on the matter content of the universe and on
cosmological parameters. In the standard inflationary scenario,
additional contributions can come from primordial gravitational waves.
Further complications could occur: the simplest models  on the origin of the 
primordial fluctuations assume adiabaticity, but other
possibilities do exist (like isocurvature modes, see Langlois, this
issue). In addition, although active perturbations, corresponding to
topological defects, are ruled out as the primary seeds of structure
formation and thereby of CMB fluctuations, the possibility remains
that a non-zero contribution does exist (Bouchet et al., 2002) which
might affect parameter estimations from the $C_\ell$'s. The increase
in precision measurements is therefore vital in order to ensures that
our vision is not blurred by such exotic contributions. Hereafter, we
will comment essentially on the interpretation of the $C_\ell$'s curve
within inflationary scenarios (see Parentani, this issue), i.e. on
passive initial Gaussian fluctuations (although non-gaussianity is
possible in inflationary scenarios).

The formalism to compute expected fluctuations in the CMB has been
developed quite early (Sachs and Wolfe, 1967; Peebles and Yu, 1970)
and useful constraints from upper limits on CMB fluctuations have been
used quite widely in the 80's (Wilson and Silk, 1981; Vittorio and
Silk, 1984; Bond and Efstathiou, 1984). However the detection of the
first fluctuations by COBE on large scale (Smoot et al; 1992) 
represents what can be considered as the most important observational
fact in Cosmology during the last twenty years of the XXth century
(although some tantalising evidence existed before DMR, there is no
doubt that the DMR instrument obtained the first reliable detection of
anisotropies beyond the dipole component).  Indeed, this discovery
lead to a deep change in modern cosmology: the DMR observations
reveals that predictions of early universe physics theories, like
inflation, were actually testable by astronomical observations. At the
same time, the DMR observations called for further effort on the
observational side: because COBE could not reveal the fluctuations on
angular scales smaller than 7 degrees, the actual information that one
can get from the DMR measurement was very limited. It has therefore
become clearer and clearer that small scale fluctuations would be
critical in bringing more stringent constraints on cosmological
scenarios. These ideas have strongly motivated the two space missions
WMAP and Planck Surveyor, as well as many balloon and ground based
experiments.

\subsection{First fundamental result: the universe is nearly flat}

\begin{figure}[!ht]
\begin{center}
\resizebox{!}{!}{\includegraphics[angle=0, height=8cm]{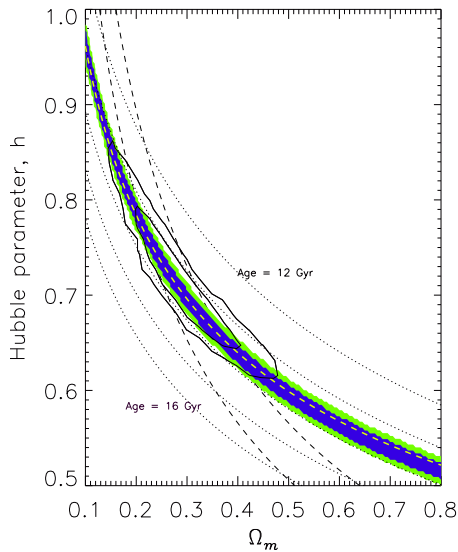}}
\end{center}
\caption{\label{figWMAP} Constraints set  by the properties 
of the first peak as seen by WMAP in the $\Omega_m - h$ plane for a
flat universe (filled region, blue, 1 $\sigma$, green 2
$\sigma$). The constraint region follows almost exactly
constant age lines (dotted). 
Additional constraints can be obtained on the dark 
matter content for powerlaw CDM
models (dashed lines), leading to tight contours when combined. 
 From Page et al. (2003).  }
\end{figure}                                         
                           
However, in order to fully exploit the result of space missions, or
even to fully explore their actual capabilities in constraining
cosmological parameters, the need for accurate and fast codes to
compute the $C_\ell$'s for large sets of parameters has become
obvious. Such extremely fast codes have become available (CMBFAST:
Seljak and Zaldarriaga, 1996; CAMB: Lewis et al, 2000; DASH:
Kaplinghat et al., 2002) allowing the computation the $C_\ell$ for a
given model in few seconds, while hours were necessary a few years
ago. Detailed investigations then become possible on a large number of
parameters (see Douspis, this issue). During the same period,
tantalising observational evidence for the presence of the first peak
was reported for the first time by the Saskatoon experiment
(Netterfield et al, 1995). Soon after, several experiments provided
measurements on similar scales. These early detections were consistent
with the presence of the so-called first Doppler peak, a maximum in
the amplitude in the $C_\ell$, found to lie close $l \sim 250$,
although their consistency was far from obvious. The first analyses of
cosmological implications revealed that this data were consistent with
flat cosmological models and inconsistent with open cosmological
models (Lineweaver et al. 1997; Handcock et al. 1998; Lineweaver and
Barbosa 1998).  During the period 1997--2000 several small scale
experiments brought further observational evidence for the detection
of the first peak as the number of measurements increased at smaller
scales (larger $\ell$).  These data consistently pointed toward a
nearly flat model, but were also pointing toward an index for the
power spectrum of initial fluctuations close to $1$ as expected in
inflationary models.  Constraints obtained from the CMB received
increasing attention, and the observational results from Boomerang and
Maxima (see Stompor et al., this issue) who provided 
maps with unprecedent S/N  brought undeniable evidence
for the presence of the Doppler peak at $\ell \sim 220$,
thereby providing the definitive evidence for a nearly flat Universe
(when data are interpreted within the framework of General Relativity;
actually the evidence for flatness-or nearly so- is not direct). This
is certainly one of the most important observational facts in modern
cosmology: the DMR result demonstrated the need for new physics, but
the present observations demonstrated without any ambiguities that
theorists had provided models whose predictions were very close to the
actual data. It is now clear that investigations of early universe
physics can be constrained -- actually quite severely -- by
astronomical observational data.

The detailed existing observations of the fluctuations of the CMB also
implied that the case for the simplest general framework, the
gravitational growth of passive Gaussian fluctuations, is very strong.
Indeed this idea is now completely accepted. It was also realised that
much tighter constraints on cosmological parameters could be obtained
by {\em combinations}: large scale structure, SNIa Hubble diagram,
Hubble constant measurements could be used in order to almost entirely
specify the value of the cosmological parameters. This technique has
been extremely fruitful with the increased accuracy of second
generation experiments (Boomerang, Maxima, CBI, Archeops, ACBAR),
see (Beno\^{\i}t et al., 2003a; Beno\^{\i}t et al., 2003b),
although early investigations did already provide crucial evidence
which lead to a standard model, the so called concordance model, now
recognised as a model able to reproduce most of existing
observations.\\

\subsection{BBN : CMB and light element abundance}

The quality of constraints that can be obtained from the $C_\ell$ is
truly remarkable. This is well illustrated by the constraints that can
be put on the baryonic content of the Universe $\Omega_b$. Somewhat
surprisingly the value of $\Omega_b$ can be constrained from available
data on the $C_\ell$, in a way which is relatively independent of the
other parameters.  Few years ago, obtaining information on this
quantity was possible only through the comparison of predictions of
primordial nucleosynthesis and the observed abundance of light
elements. Deuterium is the light element which is the most sensitive
to primordial baryon abundance. Furthermore it has now been observed
in Lyman $\alpha$ clouds which are likely not to have suffered
significant chemical evolution. A few years ago, there was a
controversy on the actual abundance of Deuterium, even before the
Boomerang data the CMB clearly favoured the lower value, indicative of
a high baryonic content. The controversy has since disappeared, and
the agreement between the baryonic content from CMB and from Deuterium
in Lyman$\alpha$ clouds is excellent (Kirkman
et al., 2003), although it is not clear whether
the abundance of Helium 4 (Gruenwald, Steigman, \& Viegas, 2003)
is fully consistent with Deuterium in standard BBN. 
It is remarkable that within a few
years the CMB has been able to provide constraints in this domain that
are of the same quality as primordial nucleosynthesis, whose reign
lasted for decades.
\\

There is no doubt that the satellites WMAP and Planck are going to
provide measurements whose accuracy will be close to the fundamental
limit implied by the so--called cosmic variance (i.e. the limited
possible knowledge on $C_\ell$ due to the finite size of the celestial
sphere). However, it is still a somewhat open question to infer what
accuracy can be achieved on Cosmological parameters. Actually, such a
question could be answered only within a specified model, and there is
some arbitrariness in deciding whether the models investigated are of
enough generality to make firm statements. For instance,
CMB data can be fit by models with $\Omega_M > 1.2$; therefore
the conclusion that CMB prefer nearly flat models relies on some a
priori. This remains a fundamental result of modern cosmology, because
the important result is that models with low $\Omega_M$ content
without cosmological constant are strongly ruled out (to my knowledge
there is no such model which could accommodate CMB data). However in
the area of {\em precision cosmology} this question deserves  special
attention. Indeed, if one writes an accurate constraint on a
cosmological parameter that CMB implies, it is wise to specify the
model in which this has been obtained. For instance, the accurate age
constraint obtained by WMAP is only meaningful within the specified
scenario (the flat power law pure CDM with adiabatic fluctuations).

\subsection{Can we be fooled?}

\begin{figure}[!ht]
\begin{center}
\resizebox{!}{!}{\includegraphics[angle=0, height=8cm]{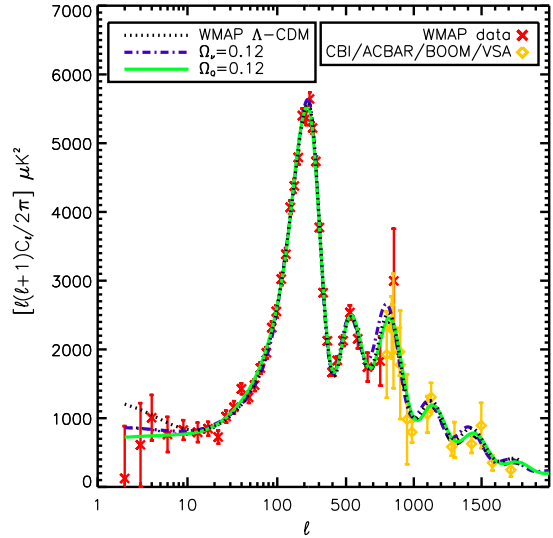}}
\end{center}
\caption{\label{figNQ} The temperature power spectrum for the best-fit
power-law $\Lambda$CDM model (dotted black line) from Spergel et al.
(2003), and for two broken-power-law models (both having $\Omega_\Lambda =
0$) with $\Omega_\nu = 0.12$ (dot-dashed blue line) and $\Omega_{\rm
\scriptscriptstyle Q}=0.12$ (solid green line), compared to data from
WMAP and other experiments (\cite{vsa,cbi,acbar,boom2}). Such models
have a low Hubble constant ($H_0 \sim 46$ km/s/Mpc) and of course are 
rejetced by the interpretation of the SNIa Hubble diagram but are consistent
with  most major cosmological data (large scale structure, abundance of
local clusters, dark matter distribution on large scales as observed
from weak lensing, primordial nucleosynthesis). From Blanchard et
al. (2003)}
\end{figure}

This question is connected to the problem of degeneracy among
cosmological parameters in estimation from the $C_\ell$. Indeed it is
well known that very different combinations of parameters could lead
to indistinguishable $C_\ell$ (Zaldarriaga et al, 1997), differences
being smaller than the cosmic variance.  In addition, current
investigations are performed assuming pure power law power spectrum
for the initial fluctuations. If there are some complexities in the
shape of the initial power spectrum may well render cosmological
constraints erroneous if this complexity is not dealt with in the
analysis (Kinney, 2001). A good example of this is the status of the
cosmological constant from the WMAP data: the detection of such term
from the CMB data is strong in pure power law $\lambda$CDM models
(Spergel et al., 2003). Allowing some type of variations in the shape
of the power spectrum leaves this conclusion essentially unchanged.
However, an Einstein de Sitter ($\Omega_M = 1$ and $\Omega_\Lambda =
0$) in which the power spectrum presented two different spectral
indexes over different scales has been shown to be able to reproduce
the WMAP results as well as the concordance model (see Blanchard et
al. 2003).

This comes from the fact that the primary cosmological quantity which
determines the $C_l$ curve is the angular distance to the last
scattering surface. Therefore, although the position of the first peak
clearly points toward a nearly flat model, there is some degeneracy
left in the $\Omega_M-H_0$ plane. This degeneracy might be easily
broken in a specific model. Indeed, the six independent parameters of
flat pure power law $\Lambda$CDM models, can be accurately determined
from the WMAP data alone. Within this framework, the emerging picture
is fully consistent with the concordance model: the index of the
primordial spectrum is very close to 1: $n = 0.99 \pm 0.04$, $H_0 = 72
\pm 5$km/s/Mpc.  Such cosmological model is also in agreement with
others measurements of cosmological relevance: the Hubble diagram of
distant SNIa, the measurement of the Hubble constant by the HST,
estimations of the matter content of the universe by various methods.
In contrast, an Einstein de Sitter model could be made consistent with
WMAP data only at the price of a low Hubble constant ($H_0 \sim
46$km/s/Mpc), which is however a value that some data would favour
(Kochanek and Schechter, 2003).

The fundamental conclusion at this level is that the concordance model
is clearly the simplest cosmological model in order to reproduce the
WMAP data.  Although one should keep in mind that formally the
WMAP data {\em rejected} the best model at more than 95\%, such a
model reach a good agreement with several data of cosmological
relevance. 

Given the importance of the hypothesis on the primordial spectrum, it
is certainly critical to have independent measurement of the power
spectrum of matter fluctuations on all scales. Surveys of galaxies as
well as the power spectrum of Lyman$\alpha$ clouds the provide such
estimation. Although they certainly provide a reasonable estimation of
the amplitude of matter fluctuations over a wide range of scales,
typically from 1 $h^{-1}$Mpc to 100 $h^{-1}$Mpc, it is much more
difficult to properly evaluate by which amount of "bias" they could be
affected.

More direct measurements of the level of fluctuations in the matter
content of the Universe, commonly expressed as $\sigma_8$, the root
mean squared amplitude over a sphere of 8 $h^{-1}$Mpc, are possible
through two techniques : the abundance of clusters and the measurement
of the weak lensing signal over large scales.  Both methods allow
rather direct measurement of a combination of $\Omega_M$ and of the
amplitude of matter fluctuations $\sigma_8$. These methods can be
extended to break the degeneracy. Both methods suffer from different
systematics which limit the present day accuracy to something like
20\% but rapid progress from large scale weak lensing surveys are
likely to allow a significant reduction of this uncertainty, allowing
a measurement of the power spectrum over a wider range of scales than
from x-ray clusters. Again a concordance model normalised to WMAP is
able to reproduce quite well the observed amplitude. 

This illustrates the remarkable success of the concordance model:
without any significant further adjustment, it is in good agreement
with what we know about large scale structure. In contrast, in an
Einstein de Sitter universe, the amplitude of matter fluctuations
derived within pure CDM models produced amplitude of matter
fluctuations on small scales which are unacceptably large.  Such a
disagreement can be alleviated by the introduction of a modest
component of matter like neutrinos or quintessence with $w \sim 0$. In
addition the power spectrum of matter fluctuations is then in
agreement  with  the observations on large scale structures.

\section{Conclusion}

The so-called concordance model provides a remarkable simple
cosmological model which reproduces well the WMAP results and which is
in agreement with a number of astrophysical observations of
cosmological relevance. Despite of this success, it should be realised
that the WMAP data by themselve do not require the introduction of a
cosmological constant. Actual direct evidence for the existence of a
non--zero cosmological constant that dominates the density of the
universe are rather limited: the Hubble diagram of distant SNIa and
the possible detection of the correlation between deep galaxy surveys
and CMB. Therefore, it is essential to confirm the actual non-zero
value of the cosmological constant (or one of its generalisation like
quintessence) by other data. The next generation of large projects
dedicated to cosmology will undoubtedly allow the reliable
establishment of a non-zero cosmological constant (if this is actually
the case...).  This will allow cosmologists to work within the robust
framework of a standard model. The high precision that should be
obtained from satellite CMB experiments, typically 1\% in the Planck
experiment (see Bouchet et al., this issue), will open the possibility
of determining the cosmological parameters with a precision of the
same order. This will be possible by combining different data that
will provide accurate, complementary information, including those on
the power spectrum of matter fluctuations. In order to take full
advantage of the accuracy of CMB data, the precision of data with
which they are combined should be similar and therefore systematic
uncertainties should be controlled with a similar precision. This is
the great challenge for precision cosmology but the rewards will be
the establishment of a standard model of cosmology to a high precision
and probably unique access to physics at energies much beyond what
would be attained directly from laboratory experiments.

%%%%%%%%%%%%%%%%%%%%%%%%%%%%%%%%%%%%%%%%%%%%%%%%%%%%%%%%%%%%
%%%  Acknowledgements  %%%
%%%%%%%%%%%%%%%%%%%%%%%%%%
\Acknowledgements{The authors would like to thank F. X. D\'esert for useful comments and corrections}
%%%%%%%%%%%%%%%%%%%%%%%%%%%%%%%%%%%%%%%%%%%%%%%%%%%%%%%%%%%%
%%%  Bibliography  %%%
%%%%%%%%%%%%%%%%%%%%%%%

%
\end{document}